\documentstyle[sprocl]{article}

\newcommand{\AmS}{{\protect\the\textfont2
  A\kern-.1667em\lower.5ex\hbox{M}\kern-.125emS}}
\def\Journal#1#2#3#4{{#1} {\bf #2}, #3 (#4)}

\def\NPB{{\em Nucl. Phys.} B}
\def\PLB{{\em Phys. Lett.}  B}
\def\PRL{\em Phys. Rev. Lett.}
\def\PRD{{\em Phys. Rev.} D}
\def\ZPC{{\em Z. Phys.} C}
\def\PTP{\em Prog. Theor. Phys.}
\newread\epsffilein 
\newif\ifepsffileok 
\newif\ifepsfbbfound 
\newif\ifepsfverbose 
\newdimen\epsfxsize 
\newdimen\epsfysize 
\newdimen\epsftsize 
\newdimen\epsfrsize 
\newdimen\epsftmp 
\newdimen\pspoints 
\def\epsfbox#1{\global\def\epsfllx{72}\global\def\epsflly{72}%
 \global\def\epsfurx{540}\global\def\epsfury{720}%
 \def\lbracket{[}\def\testit{#1}\ifx\testit\lbracket
 \let\next=\epsfgetlitbb\else\let\next=\epsfnormal\fi\next{#1}}%
\def\epsfgetlitbb#1#2 #3 #4 #5]#6{\epsfgrab #2 #3 #4 #5 .\\%
 \epsfsetgraph{#6}}%
\def\epsfnormal#1{\epsfgetbb{#1}\epsfsetgraph{#1}}%
\def\epsfgetbb#1{%
\openin\epsffilein=#1
\ifeof\epsffilein\errmessage{I couldn't open #1, will ignore it}\else
 {\epsffileoktrue \chardef\other=12
 \def\do##1{\catcode`##1=\other}\dospecials \catcode`\ =10
 \loop
 \read\epsffilein to \epsffileline
 \ifeof\epsffilein\epsffileokfalse\else
 \expandafter\epsfaux\epsffileline:. \\%
 \fi
 \ifepsffileok\repeat
 \ifepsfbbfound\else
 \ifepsfverbose\message{No bounding box comment in #1; using defaults}\fi\fi
 }\closein\epsffilein\fi}%
\def\epsfclipstring{}
\def\epsfsetgraph#1{%
 \epsfrsize=\epsfury\pspoints
 \advance\epsfrsize by-\epsflly\pspoints
 \epsftsize=\epsfurx\pspoints
 \advance\epsftsize by-\epsfllx\pspoints
 \epsfxsize\epsfsize\epsftsize\epsfrsize
 \ifnum\epsfxsize=0 \ifnum\epsfysize=0
 \epsfxsize=\epsftsize \epsfysize=\epsfrsize
 \epsfrsize=0pt
 \else\epsftmp=\epsftsize \divide\epsftmp\epsfrsize
 \epsfxsize=\epsfysize \multiply\epsfxsize\epsftmp
 \multiply\epsftmp\epsfrsize \advance\epsftsize-\epsftmp
 \epsftmp=\epsfysize
 \loop \advance\epsftsize\epsftsize \divide\epsftmp 2
 \ifnum\epsftmp>0
 \ifnum\epsftsize<\epsfrsize\else
 \advance\epsftsize-\epsfrsize \advance\epsfxsize\epsftmp \fi
 \repeat
 \epsfrsize=0pt
 \fi
 \else \ifnum\epsfysize=0
 \epsftmp=\epsfrsize \divide\epsftmp\epsftsize
 \epsfysize=\epsfxsize \multiply\epsfysize\epsftmp
 \multiply\epsftmp\epsftsize \advance\epsfrsize-\epsftmp
 \epsftmp=\epsfxsize
 \loop \advance\epsfrsize\epsfrsize \divide\epsftmp 2
 \ifnum\epsftmp>0
 \ifnum\epsfrsize<\epsftsize\else
 \advance\epsfrsize-\epsftsize \advance\epsfysize\epsftmp \fi
 \repeat
 \epsfrsize=0pt
 \else
 \epsfrsize=\epsfysize
 \fi
 \fi
 \ifepsfverbose\message{#1: width=\the\epsfxsize, height=\the\epsfysize}\fi
 \epsftmp=10\epsfxsize \divide\epsftmp\pspoints
 \vbox to\epsfysize{\vfil\hbox to\epsfxsize{%
 \ifnum\epsfrsize=0\relax
 \includegraphics{#1}%
 \else
 \epsfrsize=10\epsfysize \divide\epsfrsize\pspoints
 \includegraphics{#1}%
 \fi
 \hfil}}%
\global\epsfxsize=0pt\global\epsfysize=0pt}%
 {\catcode`\%=12 \global\let\epsfpercent=
\long\def\epsfaux#1#2:#3\\{\ifx#1\epsfpercent
 \def\testit{#2}\ifx\testit\epsfbblit
 \epsfgrab #3 . . . \\%
 \epsffileokfalse
 \global\epsfbbfoundtrue
 \fi\else\ifx#1\par\else\epsffileokfalse\fi\fi}%
\def\epsfempty{}%
\def\epsfgrab #1 #2 #3 #4 #5\\{%
\global\def\epsfllx{#1}\ifx\epsfllx\epsfempty
 \epsfgrab #2 #3 #4 #5 .\\\else
 \global\def\epsflly{#2}%
 \global\def\epsfurx{#3}\global\def\epsfury{#4}\fi}%
\def\epsfsize#1#2{\epsfxsize}
\def\sss{\scriptscriptstyle}
\def\barp{{\raise.35ex\hbox{${\sss (}$}}---{\raise.35ex\hbox{${\sss )}$}}}
\def\bdbarp{\hbox{$B_d$\kern-1.4em\raise1.4ex\hbox{\barp}}}
\def\bsbarp{\hbox{$B_s$\kern-1.4em\raise1.4ex\hbox{\barp}}}
\def\dbarp{\hbox{$D$\kern-1.1em\raise1.4ex\hbox{\barp}}}

\newcommand{\ra}{\rightarrow}

\def\be{\begin{equation}}
\def\ee{\end{equation}}
\def\g{\gamma}

\def\mt{m_t}
\def\mb{m_b}
\def\mc{m_c}

\def\bea{\begin{eqnarray}}
\def\eea{\end{eqnarray}}
\def\be{\begin{equation}}
\def\ee{\end{equation}}

\def\g{\gamma}

\def\mt{m_t}

\newcommand{\bgamaxs}{$B \to X _{s} + \gamma$}

\newcommand{\BGAMAXS}{B \ra X _{s} + \gamma}

\newcommand{\BBGAMAXS}{{\cal B}(B \ra  X _{s} + \gamma)}
\newcommand{\BBGAMAXD}{{\cal B}(B \ra  X _{d} + \gamma)}

\newcommand{\BGAMAKSTAR}{B \ra  K^{\star} + \gamma}

\def\beq{\begin{equation}}
\def\eeq{\end{equation}}

\def\Vcbabs{\vert V_{cb} \vert}

\def\Vtdabs{\vert V_{td} \vert}

\def\Vtsabs{\vert V_{ts} \vert}

\newcommand{\absvcb}{\vert V_{cb}\vert}

\newcommand{\absvts}{\vert V_{ts}\vert}
\newcommand{\absvtb}{\vert V_{tb}\vert}

\newcommand{\fbb}{f^2_{B_d}B_{B_d}}
\newcommand{\fbbs}{f^2_{B_s}B_{B_s}}
\newcommand{\fbd}{f_{B_d}}

\newcommand{\delmd}{\Delta M_d}
\newcommand{\delms}{\Delta M_s}

\def\qbar{\overline q}

\def\q5q{\qbar{{\lambda_a}\over 2} i\gamma_5 q}

\newcommand{\bgamaxd}{$B \to X _{d} + \gamma$}

\def\to{\rightarrow}

\def\mb{m_b}

\def\xs{x_s}

\newcommand{\bdbdbar}{$B_d^0$-${\overline{B_d^0}}$}
\newcommand{\bsbsbar}{$B_s^0$-${\overline{B_s^0}}$}

\def\as{\alpha _s}

\begin{document}
\begin{flushright}
DESY 97-256\\
December 1997\\
\end{flushright}
\begin{center}
{\Large \bf
\centerline{FLAVOUR CHANGING NEUTRAL CURRENT PROCESSES}
\vspace*{0.2cm}
\centerline{AND CKM PHENOMENOLOGY}}
\vspace*{1.5cm}
 {\large A.~Ali}
\vskip0.2cm
 Deutsches Elektronen-Synchrotron DESY, Hamburg \\
Notkestra\ss e 85, D-22603 Hamburg, FRG\\

\vspace*{8.0cm} 
To be published in the Proceedings of the First APCTP Workshop\\
Pacific Particle Physics Phenomenology,
Oct.~31 - Nov.~2, 1997,\\
Seoul National University, Seoul, South Korea

\end{center}
\thispagestyle{empty}
\newpage   
\setcounter{page}{1}
\title{FLAVOUR CHANGING NEUTRAL CURRENT PROCESSES AND CKM
PHENOMENOLOGY}

\author{ A. Ali}

\address{Deutsches~Elektronen-Synchrotron~DESY,
     Notkestrasse~85,~D-22603~Hamburg,~FRG}

%


\maketitle\abstracts{
Some selected topics in flavour changing neutral current processes in
$B$ decays are discussed in the context of the Standard Model. The
emphasis here is on the determination of the CKM matrix elements from
the ongoing and forthcoming experiments. The topics discussed include:
(i) CKM-allowed radiative decays $B \to X_s + \gamma$, 
 (ii) CKM-suppressed radiative decays $B \to X_d + \gamma$, $B^\pm \to 
\rho^\pm + \gamma$ and $B^0 \to (\rho^0,\omega) + \gamma$,
(iii) $B^0$ - $\overline{B^0}$-mixings, and
(iv) the Fleischer-Mannel bound in $B \to K\pi$ decays.}
%
%
\section{Determination of $\Vtsabs$ from $\BBGAMAXS$}
\subsection{Experimental status}
Electromagnetic penguins in $B$ decays were first measured in 1993 
by the CLEO collaboration through the exclusive
decay $\BGAMAKSTAR$ \cite{CLEOrare1},
followed by the measurement of the inclusive 
decay $\BGAMAXS$ in 1994 \cite{CLEOrare2}.
The present CLEO measurements can be summarized as \cite{Tomasz97}: 
\begin{eqnarray}  
\label{penguinexp}
{\cal B}(\BGAMAXS) &=& (2.32\pm 0.57\pm 0.35)\times 10^{-4}, \nonumber\\
{\cal B}(\BGAMAKSTAR) &=& (4.2\pm 0.8 \pm 0.6)\times 10^{-5},
\end{eqnarray}
which yield an exclusive-to-inclusive ratio:
$R_{K^*} \equiv \Gamma(\BGAMAKSTAR)/\Gamma(\BGAMAXS)=(18.1\pm
6.8)\% $.
Very recently, the
inclusive radiative decay has also been reported by the ALEPH collaboration 
with a (preliminary) branching ratio \cite{ALEPHbsg}: 
\begin{equation}
\label{alephbsg}
{\cal B}(H_b \to X_s + \gamma) = (3.29 \pm 0.71 \pm 0.68)\times 10^{-4}. 
\end{equation}
 The  branching ratio in (\ref{alephbsg}) involves a 
different weighted average of the various $B$-mesons and $\Lambda_b$ baryons
produced in $Z^0$ decays (hence the symbol $H_b$ ) than the 
corresponding one
given in (\ref{penguinexp}), which has been measured in the decay 
$\Upsilon (4S) \to B^+ B^-, B^0 \overline{B^0}$.
 
In the context of SM, the principal interest in the inclusive branching 
ratio $\BGAMAXS$ lies in that it determines the 
ratio of the Cabibbo-Kobayashi-Maskawa (CKM) matrix elements  $\vert 
V_{ts}^* V_{tb}/V_{cb}\vert$.
Since $\absvcb$ and $\absvtb$ have been directly measured,  
the measurements (\ref{penguinexp}) and (\ref{alephbsg}) yield  
$\absvts$. The ratio $R_{K^*}$ constrains the
form factor in $B \to K^* + \gamma$. Estimates of this ratio in the
QCD sum rule and lattice QCD \cite{Alisb97} are in agreement with the 
present measurements.

\subsection{SM estimates of ${\cal B}(\BGAMAXS)$ and ${\cal B}(H_b \to 
X_s + \gamma)$}
The leading contribution to the decay $b \to s +\gamma$ arises
at one-loop from the so-called penguin diagrams. With the help of the
unitarity of the CKM matrix,
the decay matrix element in the lowest order can be written as:
\be
\label{e2}
{\cal M }(b \to s ~+\gamma)
    = \frac{G_F}{\sqrt{2}}\,\frac{e}{2 \pi^2} \,\lambda_{t}
(F_2 (x_t)-F_2(x_c))
 q^\mu \epsilon^\nu \bar{s} \sigma_{\mu \nu} (m_bR ~+ ~m_sL)b ~.
\ee
where $G_F$ is the Fermi coupling constant, $e=\sqrt{4 \pi 
\alpha_{\mbox{em}}}$, $x_i= ~m_i^2/m_W^2; ~i=u,c,t$ are the scaled quark 
mass ratios, and 
$q_\mu$  and $\epsilon_\mu$ are, respectively, the photon four-momentum
and polarization vector. Here, $F_2(x_i)$ is the Inami-Lim function 
derived from the (1-loop)
penguin diagrams \cite{InamiLim} and the
CKM-matrix element dependence is factorized in $\lambda_t\equiv V_{tb} 
V_{ts}^*$.
As the inclusive decay widths of the $B$ hadrons are proportional to
$\absvcb^2$, 
 the measurement of  $\BBGAMAXS$ can be 
readily interpreted in terms of the CKM-matrix element ratio
$\lambda_t/\Vcbabs$.
The required QCD radiative
and power corrections have been computed in the effective Hamiltonian
approach up to next-to-leading order (NLO) logarithmic accuracy and to 
leading order in $1/m_b^2$ and $1/m_c^2$.
The effective Hamiltonian in the SM can be written as (keeping operators 
up to dimension 6),
 \begin{equation}\label{heffbsg}
{\cal H}_{eff}(b \to s +\gamma) = - \frac{4 G_F}{\sqrt{2}} \lambda_t
        \sum_{i=1}^{8} C_i (\mu) {\cal O}_i (\mu) ,
\end{equation}
where the operator basis, the lowest order coefficients
$C_{i}(m_W)$ and the renormalized coefficients $C_{i}(\mu)$ can be
seen elsewhere \cite{ALI96}.

 Expressing the
 branching ratio  $\BBGAMAXS$ in terms of the
semileptonic decay branching ratio ${\cal B} (B \to X\ell \nu_\ell)$,
\begin{equation}
\label{brbsgsm}
{\cal B} ( B \ra  X_{s} \g) = [\frac{\Gamma(B \ra  
\gamma + X_{s})}{\Gamma_{SL}}]^{th}
\, {\cal B} (B \to X\ell \nu_\ell), 
\end{equation}  
the theoretical part denoted by $[...]^{th}$ has been calculated in the NLO 
accuracy \cite{Misiak96,Greub97,BKP97}.
The power corrections have been obtained assuming that the
decay $\BGAMAXS$ is dominated by the magnetic moment operator $O_7$.
The calculated $1/m_b^2$ correction is found to be innocuous 
(contributing at about $1\%)$ in $\BBGAMAXS$.
The correction proportional to $1/m_c^2$, resulting from the
interference of the operators $O_2$ and $O_7$ in $H_{eff}(b \to s \gamma)$, 
has also been worked out \cite{Voloshinbsg,powermc,BIR97}. 
Expressing this symbolically as 
\begin{equation}
\frac{\Gamma(\BGAMAXS)}{\Gamma^{0}(\BGAMAXS)} = 1 + \frac{\delta_c}{m_c^2},
\end{equation} 
one finds \cite{BIR97} $\delta_c/m_c^2 \simeq + 0.03$.

Using $|V_{ts}^* V_{tb}/V_{cb}|=0.976 \pm 0.010$, obtained from the 
CKM-unitarity \cite{PDG96}, and 
the current measurements  
\cite{Feindt97}:  ${\cal B}(B \to X \ell \nu_\ell) =(10.49 \pm 0.46)\%$
(at $\Upsilon(4S))$ and ${\cal B}(H_b \to X \ell \nu_\ell) =(11.16 \pm 
0.20)\%$ (at $Z^0$),
 yields
\begin{eqnarray}
\label{upssmali}
{\cal B} (\BGAMAXS ) &=& (3.51 \pm 0.32) \times 10^{-4}~,\\ \nonumber
{\cal B} (H_b \to X_s + \gamma ) &=& (3.76 \pm 0.30) \times 10^{-4}~,
\end{eqnarray}
to be compared with the CLEO measurement ${\cal B} (\BGAMAXS )= (2.32 \pm 
0.67) \times 10^{-4}$ and the ALEPH measurement
${\cal B} (H_b \to X_s + \gamma )= (3.29 \pm 0.98) \times 10^{-4}$,
respectively.

Letting the CKM factor $|V_{ts}^* V_{tb}/V_{cb}|$ as a free parameter, the 
NLO SM-based theory and experiments have been used to  
determine the CKM factor, yielding \cite{Alisb97}
\begin{equation}
|\frac{V_{ts}^* V_{tb}}{V_{cb}}| = 0.84  \pm  0.10 ~.
\label{vcs}
\end{equation}
 With the CKM
unitarity, one has $ |\frac{V_{ts}^* V_{tb}}{V_{cb}}| \simeq |V_{cs}|$;
this equality holds numerically (within present precision) if one compares 
eq.~(\ref{vcs}) 
with $|V_{cs}|= 1.01 \pm 0.18$, determined from charmed hadron decays 
\cite{PDG96}. Using the value of $\absvtb$
measured  by the CDF collaboration \cite{CDFvtb}, $\vert V_{tb} 
\vert = 0.99 \pm 0.15$, and noting  that $\absvcb = 0.0393 
\pm 0.0028$ \cite{Gibbons96}, finally yields
 \begin{equation}
\absvts = 0.033 \pm 0.007~.
\end{equation}
This is probably as direct a determination of $\vert V_{ts}\vert$ as we will 
ever see, as the  decay $t \to W +s$ is too daunting to measure due
to the low tagging efficiency of the $s$-quark jet.

%
%
\section{The decay \bgamaxd\ and constraints on the CKM parameters} %
\indent
 In close analogy
with the \bgamaxs\ case,
the complete set of dimension-6 operators relevant for
the processes $b \to d \gamma$ and $b \to d \gamma g$ 
can be written as:
\begin{equation}
\label{heffd}
{\cal H}_{eff}(b \to d)=
 - \frac{4 G_{F}}{\sqrt{2}} \, \xi_{t} \, \sum_{j=1}^{8}
C_{j}(\mu) \, \hat{O}_{j}(\mu),\quad
\end{equation}
where $\xi_{j} = V_{jb} \, V_{jd}^{*}$ with $j=u,c,t$. The 
current-current operators
 $\hat{O}_j, ~j=1,2$, have implicit in them CKM factors $\xi_c$ and $\xi_u$.
We shall use the Wolfenstein parametrization \cite{Wolfenstein},   
in which case the matrix is determined in terms of the four parameters
$A, \lambda=\sin \theta_C$, $\rho$ and $\eta$, and one can express the above
factors as :
\begin{equation} 
\xi_u = A \, \lambda^3 \, (\rho - i \eta),
~~~\xi_c = - A \, \lambda^3 ,
~~~\xi_t=-\xi_u - \xi_c.
\end{equation}
We note that all three CKM-angle-dependent quantities
$\xi_j$ are of the
same order of magnitude, $O(\lambda^3)$.
The branching ratio  $\BBGAMAXD$ in the SM  can be quite generally 
written as:
\bea
\label{branstruc}
\lefteqn{\BBGAMAXD = D_1 \lambda^2}
\nonumber\\&&{}
 \{(1-\rho)^2 + \eta^2 -(1-\rho) D_2 - \eta D_3 +D_4  \} , \quad
\eea
where the functions $D_i$ depend on various parameters such 
as $\mt,\mb,\mc,\mu$, and $\as$.
These functions were calculated in the LL approximation some time ago 
\cite{ag2} and since then their estimates have been improved 
\cite{aag97}, transcribing the NLO calculations done for $B \to X_s + 
\gamma$, thereby reducing the scale dependence in $\BBGAMAXD$ which is 
marked in the LL result.

 To get an estimate of  
$\BBGAMAXD$, the CKM parameters $\rho$ and $\eta$ have 
to be constrained from the unitarity fits \cite{AL96}.
Allowing these parameters to vary over the entire allowed domain, one
gets (at 95\% C.L.) \cite{aag97}:
\begin{equation}
 6.0 \times 10^{-6} \leq \BBGAMAXD \leq 3.0 \times 10^{-5}.
\end{equation}
The present theoretical uncertainty in this rate is a factor 5, almost all
of which is due to the CKM parameters,
 which shows that even a modest (say, $\pm 30\%)$
measurement of $\BBGAMAXD$ will have a very significant impact on the
CKM phenomenology. The decay $B \to X_d +\gamma$  is of interest from the 
point of view of direct CP 
violation as well. The CP asymmetry in this mode is estimated as 
$O(10\%)$.
 This should also reflect itself in the rate asymmetry in some 
exclusive decays such as $B^\pm \to \rho^\pm + \gamma$.
\section{CKM-suppressed exclusive radiative decays $B \to V + \gamma$}
 \par
Exclusive radiative
 $B$ decays $B \to V + \gamma$, with $V=K^*,\rho,\omega$, are also 
potentially very interesting for the CKM phenomenology
\cite{abs93}. Extraction of CKM parameters from these decays would, however, 
involve a trustworthy estimate of the (short distance) SD- and 
(long distance) LD-contributions in the decay amplitudes.
Present estimates based on QCD sum rules \cite{wyler95,ab95} and 
soft-scattering models \cite{dgp97} yield small ($O(5\%)$ to moderate 
$(O(20\%)$ LD-contributions in these decays depending on the decay modes.
More importantly, data on neutral and charged $B$ mesons can be used to 
disentangle the LD-effects directly.  
 \par
  The SD-contribution in the 
 exclusive decays $(B^\pm, B^{0}) \to (K^{*\pm}, K^{* 0})+ \gamma$,
$(B^\pm, B^{0}) \to (\rho^\pm,\rho^{0}) + \gamma$,
$B^{0} \to \omega + \gamma$  and the
corresponding $B_s$ decays, $B_s \to \phi + \gamma $, and
$B_s \to K^{* 0} + \gamma $,
involve the magnetic moment operator ${\cal O}_7$ and the related one 
obtained by the change $s \to d$, $\hat{O}_7$.
Keeping only the SD-contribution leads to obvious relations among the
exclusive decays, exemplified here by the decay
rates for $(B^\pm,B^0) \to (\rho^\pm,\rho^0) + \gamma$ and $(B^\pm,B^0) \to 
(K^{*\pm},K^{*0}) + \gamma$: 
\be
\frac{\Gamma ((B^\pm,B^{0}) \to (\rho^\pm,\rho^{0}) + \gamma)}
     {\Gamma ((B^\pm,B^{0}) \to (K^{*\pm},K^{* 0}) + \gamma)} 
  \simeq \kappa_{u,d}\left[\frac{\Vtdabs}{\Vtsabs}\right]^2 \,,
\label{SMKR}
\ee
where $\kappa_{i} \equiv [F_S(B_i \to \rho \gamma)/F_S(B_i \to K^* 
\gamma)]^2$ involves the ratio of the transition form factors
(apart from the small phase-space dependent correction) in the 
indicated radiative $B$ decays; this
ratio is unity in the $SU(3)$ limit. Likewise, assuming 
dominance of SD physics gives relations among various decay rates
 \beq\label{ratio2}
\Gamma(B^\pm \to \rho^\pm \gamma)=2 ~\Gamma(B^{0}\to \rho^0  \gamma)
    = 2 ~\Gamma (B^{0} \to \omega  \gamma)~,
\eeq
where the first equality holds due to the isospin invariance,
and in the second $SU(3)$ symmetry has been assumed.

The LD-amplitudes in radiative $B$ decays from the light quark 
intermediate states necessarily involve CKM matrix elements
other than $V_{td}$ or $V_{ts}$. 
In the CKM-suppressed decays $B \to V + \gamma$ the LD-contributions are
dominantly induced by the matrix elements of the
four-Fermion operators $\hat{O}_1$ and $\hat{O}_2$. 
Using factorization, the LD-amplitude in the decay $B^\pm \to \rho^\pm + 
\gamma$ can be written in terms of two form factors $F_1^L$ and $F_2^L$.
These are found to be  numerically close to each other in the 
QCD sum rule approach \cite{wyler95,ab95}, $F_1^L\simeq F^L_2 \equiv F_L$,
hence the ratio of the LD- and the SD- contributions reduces to a number 
 \begin{equation}\label{ratio2p}
{\cal A}_{long}/{\cal A}_{short}=
R_{L/S}^{B^\pm\to\rho^\pm\gamma}
\cdot\frac{V_{ub}V_{ud}^\ast}{V_{tb}V_{td}^\ast} ~.
\end{equation}
where $R_{L/S}^{B^\pm\to\rho^\pm\gamma}$ is estimated as \cite{ab95}: 
\be
\label{result2}
R_{L/S}^{B^\pm\to\rho^\pm\gamma}  = -0.30\pm 0.07 ~.
\ee
The CKM ratio in Eq.~(\ref{ratio2p}) is constrained by unitarity to lie in
the range \cite{AL96} $0.22 \leq V_{ub}V_{ud}^\ast/V_{tb}V_{td}^\ast \leq 
0.71$, with the central value around $0.33$. This would put the 
LD-corrections at $O(10 -20)\%$. The analogous LD-contributions to the 
neutral $B$ decays $B^{0}\to\rho\gamma $ and $B^{0}\to\omega\gamma $ are
expected to be much smaller due to the electric charge and QCD colour 
factors, and are estimated to be \cite{ab95,dgp97}
$R_{L/S}^{B^{0}\to\rho\gamma} \simeq R_{L/S}^{B^{0}\to\omega\gamma}=0.05$.

 The relations (\ref{ratio2}), which hold on ignoring LD-contributions,
get modified by including the LD-contributions.
Likewise, the ratios of the CKM-suppressed and CKM-allowed
decay rates given in eq.~(\ref{SMKR}) get  modified due to the LD 
contributions.
The effect of the LD-contributions is estimated to be modest but not 
negligible in charged $B$ decays, introducing an uncertainty of $O(15\%)$,  
comparable to the  uncertainty in the overall normalization
due to the $SU(3)$-breaking effects in the quantity $\kappa_u$.
Neutral $B$-meson radiative decays are less-prone to the LD-effects,
and hence one expects that to a good approximation
(say, better than $10\%$) the ratio of the decay rates for neutral $B$ meson 
obtained in the approximation of SD-dominance remains valid \cite{abs93}:
\begin{equation}
\frac{\Gamma(B^0\to \rho\gamma,\omega\gamma)}{\Gamma(B\to K^*\gamma)}
 = \kappa_d\lambda^2 [(1-\rho)^2+\eta^2]~,
\end{equation}
where this relation holds for each of the two decay modes separately.

 Finally, combining the estimates for the LD- and SD-form factors
\cite{ab95,abs93},  and restricting the Wolfenstein
parameters in the allowed range given earlier, yields:
${\cal B}(B^\pm\to \rho^\pm\gamma)
= (1.5 \pm 1.1) \times 10^{-6} ~,
{\cal B}(B^{0}\to \rho\gamma) \simeq {\cal B}(B^{0}\to \omega \gamma)
=  (0.65 \pm 0.35) \times 10^{-6}$.
 The large range reflects to a large extent the poor
knowledge of the CKM matrix elements and hence experimental measurements
of these branching ratios will contribute greatly to determine the
Wolfenstein parameter $\rho$ and $\eta$.
 Present experimental limits 
(at $90\%$ C.L.) are \cite{Tomasz97}:
${\cal B}(B^\pm\to \rho^\pm\gamma) <1.1 \times 10^{-5}$,
${\cal B}(B^{0}\to \rho\gamma) < 3.9 \times 10^{-5}$ and
${\cal B}(B^{0}\to \omega \gamma) < 1.3 \times 10^{-5}$. The constraints
on the parameters $(\rho,\eta)$ following from them are, however, not yet
competitive to the ones which follow from the CKM-unitarity and lower 
bound on the mass difference ratio $\delms/\delmd$ in the
 $B_s^0$ - $\overline{B_s^0}$ and $B_d^0$ - $\overline{B_d^0}$
 sectors, which we discuss next. 

\section{$\delms$ (and $\xs$), Unitarity Triangle and CP Phases}

Mass differences in the \bsbsbar\ and \bdbdbar\
systems are dominated in the standard model by the $t$-quark exchange.
Of the two, $\delmd$ is well-measured \cite{Feindt97}, $\delmd =(0.464 \pm 
0.018)(\mbox{ps})^{-1}$,
which can be translated into a determination of the matrix element $\vert 
V_{td}\vert$, yielding \cite{AL96}
$\vert V_{td}\vert = (9.0 \pm 2.6)\times 10^{-3}$, where the error reflects 
dominantly theoretical uncertainty on the hadronic quantity, for which 
$f_{B_d}\sqrt{\hat{B}_{B_d}}=(200 \pm 40)$ MeV has been used. The mass 
difference
$\delms$ depends quadratically on the CKM factor $\vert V_{ts}^* V_{tb} 
\vert$, which has been measured from the branching ratio $B \to 
X_s +\gamma$, yielding
$\vert V_{ts}^* V_{tb} \vert = 0.033 \pm 0.007$.
However, as the error on this quantity is more than a factor 2 larger than 
what 
one gets from the CKM unitarity, we use here the value obtained by unitarity,
$\vert V_{ts}^* V_{tb} \vert = 0.038 \pm 0.0027$. This yields:
\begin{eqnarray}
\delms &=& \left(12.0 \pm 2.0\right)\frac{\fbbs}{(230~\mbox{MeV})^2}
~(\mbox{ps})^{-1}~, \nonumber \\
\xs &=& \left(18.2 \pm 3.0\right)\frac{\fbbs}{(230~\mbox{MeV})^2}~,
\end{eqnarray}
where we have used $\tau(B_s)=(1.52 \pm 0.07)$ ps to calculate $x_s$.
The choice $f_{B_s}\sqrt{\hat{B}_{B_s}}= 230$ MeV corresponds to the 
central value given by the lattice-QCD estimates. Allowing the
coefficient to vary by $\pm 2\sigma$, and taking the central value for
$f_{B_s}\sqrt{\hat{B}_{B_s}}$,  gives
\begin{eqnarray}
8.0 ~(\mbox{ps})^{-1} &\leq & \delms \leq 16.0 ~(\mbox{ps})^{-1}~, 
\nonumber\\
 12.2 &\leq & \xs \leq 24.2~.
 \label{bestxs}
\end{eqnarray}
It is difficult to ascribe a confidence level to this range due to the
dependence on the unknown coupling constant factor.
The present lower bound (from LEP) on $\delms$ is: $\delms > 
10.0~(\mbox{ps})^{-1}$ (95\% C.L.) \cite{Moser97}.
A useful quantity for the CKM phenomenology is the ratio
\beq
\frac{\delms}{\delmd} =
 \frac{\hat{\eta}_{B_s}M_{B_s}\left(\fbbs\right)}
{\hat{\eta}_{B_d}M_{B_d}\left(\fbb\right)}
\left\vert \frac{V_{ts}}{V_{td}} \right\vert^2~,
\label{xratio}
\eeq
in which all dependence on the $t$-quark mass drops out, 
leaving the square 
of the ratio of CKM matrix elements, multiplied by a factor which reflects
$SU(3)_{\rm flavour}$-breaking effects.

\begin{figure}
\vskip -1.0truein
\centerline{\epsfxsize 3.5 truein \epsfbox {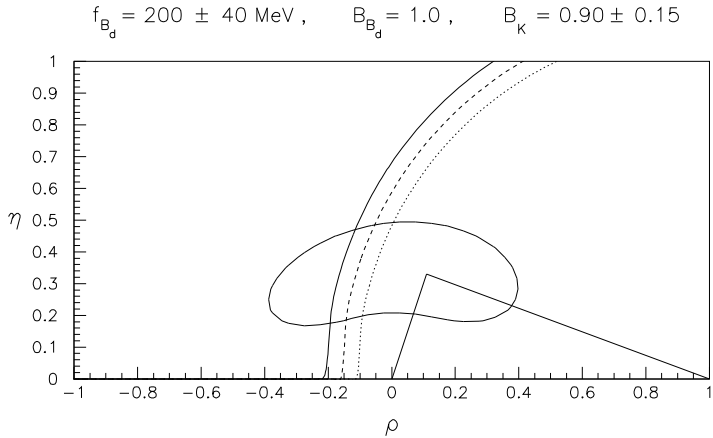}}
\vskip -1.4truein
\caption{Constraints in $\rho$-$\eta$ space from the LEP bound
 $\delms/\delmd > 20.4$. The bounds are presented for 3 choices of
the SU(3)-breaking
parameter: $\xi_s^2 = 1.21$ (dotted line), $1.32$ (dashed line) and $1.44$
(solid line). In all cases, the region to the left of the curve is ruled
out. (Updated from ref.~21.)}
\label{xslimit}
\end{figure}

The present lower bound on $\delms$ from LEP and the world average of 
$\delmd$ 
 can be used to put a bound on the ratio
$\delms/\delmd$, yielding $\delms/\delmd \geq 20.4$ at 95\% C.L. This is 
significantly better than the lower bound on this quantity from the CKM
 fits \cite{AL96}:
 $\delms/\delmd \geq 11.4 ~(\xi_s/1.15)^2$,
where $\xi_s \equiv (f_{B_s} \sqrt{\hat{B}_{B_s}}) / (f_{B_d}
\sqrt{\hat{B}_{B_d}})$ is theoretically estimated to be $\xi_s = (1.15 \pm
0.05)$.
The 95\% confidence limit on
$\delms/\delmd$ can be turned into a bound on the CKM parameter space
$(\rho,\eta)$ by choosing a value for the SU(3)-breaking parameter
$\xi_s$. We assume three representative values: $\xi_s^2 = 1.21$, $1.32$
and $1.44$, and display the resulting constraints in Fig.~\ref{xslimit}.
From this graph we see that the LEP bound now restricts the allowed
$\rho$-$\eta$ region for all three values of $\xi_s^2$. We note that the
bound on $\delms/\delmd$ removes a large (otherwise allowed) 
negative-$\rho$ region. The resulting fits of the CKM triangle
taking into account this bound are necessarily asymmetric around
$\rho=0$. The best fit solutions are around $\rho=0.11, \eta=0.33$,
as also noted by Paganini et al. \cite{paganini97}.

\begin{figure}
\vskip -0.5truein
\centerline{\epsfxsize 2.5 truein \epsfbox {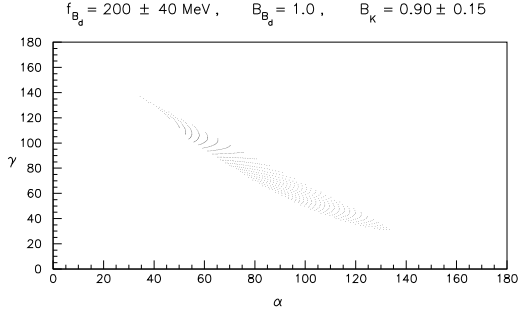}}
\vskip -1.2truein
\caption{Allowed values (in degrees) of the angle $\alpha$ and $\gamma$
at $95\%$ C.L. resulting from the CKM fits taking into account the LEP bound
$\delms/\delmd > 20.4$, with the indicated ranges of
$f_{B_d}\sqrt{\hat{B}_{B_d}}$ and $B_K$. (Updated from ref.~21.)} 
 \label{alphagam}
\end{figure}

The CP-violating asymmetries in $B$ decays can be expressed 
straightforwardly in terms
of the CKM parameters $\rho$ and $\eta$. The 95\% C.L.\ constraints on
$\rho$ and $\eta$ just discussed can be used to predict the 
correlated ranges of the angles of the CKM unitarity triangle $\alpha$, 
$\beta$ and $\gamma$ in the standard model.
These correlations were worked out earlier without the $\delms/\delmd$
bound \cite{AL96,ALI96} and are being updated here. It is found that
the effect of the LEP bound on $\delms/\delmd$ is still marginal on the
$\sin 2 \beta$ - $\sin 2 \alpha$ correlation; a small region in $\sin 2 
\beta$  below $\sin 2 \beta =0.3$ is now removed, with the (95\% C.L.) range
for $\sin 2 \alpha$ remaining 
practically unconstrained with or without the LEP bound. The correlation in
the angles $\alpha$ - $\gamma$ is more susceptible to the aforementioned 
bound. The allowed range for this correlation also 
depends more sensitively on the value of the quantity $B_K$ than is the
case for the $\sin 2 \beta$ - $\sin 2 \alpha$ correlation. 
 The previously allowed range \cite{AL96,ALI96} (at 95\% 
C.L.): $\gamma = (90\pm 50)^\circ$ 
was based on using the bag parameter to have a value in the range $B_K=0.9 
\pm 0.1$. Taking a somewhat larger 
error on the advice of our lattice colleagues, $B_K=0.9 \pm 0.15$, this 
range becomes larger. One now gets 
 $\gamma = (90\pm 58)^\circ$ without the LEP bound. The corresponding 
(95 \% C.L.) range with the LEP bound is:
 $32^\circ \leq \gamma \leq 120^\circ$. Thus, values of the angle 
$\gamma$ in excess of $120^\circ$ are now excluded by the LEP bound on
$\delms/\delmd$. Moreover,
the allowed  range in $\gamma$ is no longer symmetric around $\gamma 
=90^\circ$. This is illustrated  in Fig.~\ref{alphagam},
which shows the region in
$\alpha$-$\gamma$ space allowed by the present data, for 
$\fbd\sqrt{\hat{B}_{B_d}}=200 \pm 40$ MeV and $B_K=0.9\pm 0.15$.

 The ranges for the CP-violating rate asymmetries parametrized by $\sin
2\alpha$, $\sin 2\beta$ and and $\sin^2 \gamma$ are determined at 95\% C.L.
to be:
\begin{eqnarray}
&~& -1.0 \leq \sin 2\alpha \le 1.0~, \nonumber \\
&~& 0.30 \leq \sin 2\beta \le 0.88~, \\
&~& 0.27 \leq \sin^2 \gamma \le 1.0~. \nonumber
\end{eqnarray}
The CP asymmetries just discussed will be measured through rate
asymmetries in non-leptonic $B$ decays, such as
 $\bdbarp \to J/\psi K_S$ ($\sin 2 \beta)$, $\bdbarp \to \pi^+
\pi^-$ ($\sin 2 \alpha)$ and $\bsbarp\ \to D_s^\pm K^\mp$ (or $B^\pm \to
\dbarp\ K^\pm$) ($\sin ^2 \gamma)$. 
\section{The Fleischer-Mannel bound in $B \to K\pi$ decays}
 Recently, the role of the decays $B^0 \to \pi^\mp K^\pm$ 
and $B^\pm \to \pi^\pm K^0$ in determining the angle $\gamma$ has been 
emphasized by Fleischer and Mannel \cite{FM97}. Assuming that the 
so-called Tree (current-current) and QCD penguins are the dominant 
contributions in these decays, one can write the amplitudes as
\begin{eqnarray*}
\Gamma(B^0 \to \pi^\mp K^\pm) &\propto & |A_{P}|^2(1- 2r \cos \gamma \cos
\delta + r^2)~, \nonumber\\
\Gamma(B^\pm \to \pi^\pm K^0) &\propto & |A_{P}|^2~, \nonumber
\end{eqnarray*}
 where $r= |A_T|/|A_P|$ is the ratio of the Tree and Penguin amplitudes, 
and  $\delta$ is the strong phase difference
involving these amplitudes. This leads to the 
Fleischer-Mannel relation (an average over CP-conjugate modes is
implied), 
\begin{equation}
 R \equiv \frac{\Gamma(B^0 \to 
\pi^\mp K^\pm)}{\Gamma(B^\pm \to
\pi^\pm K^0)} = 1-2\, r \cos \gamma \cos \delta + r^2 \geq \sin^2 \gamma~.
\end{equation}
From this, constraints on $\gamma$ of the form
\begin{equation}
 0^\circ \leq \gamma \leq \gamma_0 ~~\vee ~~ 180^\circ - \gamma_0 \leq \gamma
\leq 180^\circ ~
\end{equation}
follow, where $\gamma_0$ is the maximum value of 
$\gamma$, which
are complementary to the ones from the CKM unitarity fits discussed above. 
 For $R < 1$, a useful bound on $\gamma$ emerges,
$\gamma_o^{\mbox{max}} = \cos ^{-1} (\sqrt{1-R})$.
At present $R$ has the value \cite{CLEOrare3}: $R= 0.65 \pm 0.40$. 
 For the central 
value of $R$, for example, this yields $\gamma_o^{\mbox{max}}=54^\circ$,
implying 
$ 0^\circ \leq \gamma \leq 54^\circ ~\vee ~126^\circ \leq \gamma \leq
180^\circ$. One then finds that there is no overlap in the range $\gamma 
> 90^0$ (corresponding to $\rho < 0$ in Fig.~\ref{xslimit}) with the 
allowed range from the CKM fits, but there is
significant overlap in the range $\gamma < 90^\circ$ (or $\rho > 0$ 
in Fig.~\ref{xslimit}), yielding $32^\circ \leq \gamma \leq 
54^\circ$ for the assumed value of $R$. However, as $R$ increases, the 
overlap with the parameter space allowed by the 
CKM-unitarity increases. Hence, with the present experimental error on
$R$, no useful bound on $\gamma$ follows: For the Fleischer-Mannel bound to
be useful the ratio $R$ must be significantly less 
than 1. With model-dependent estimates of $r$ and $\delta$, such as 
those obtained in the context of
factorization models, sharper constraints follow on the CKM parameters. For 
example,
$r$ is estimated to lie in the range $0.15 \leq r \leq 0.30$ in such
models, one then finds \cite{AG97} that 
$ R= 0.65 \pm 0.40$ implies $\rho \geq 0 $ (at $\pm 1 \sigma$). 
Possible dilution of the Fleischer-Mannel bounds from
the neglected contributions (such as annihilation, electroweak penguins 
and soft rescattering) have been anticipated \cite{Neubert97}. Some of these
contributions may introduce significant uncertainties in the extraction
of $\gamma$ from this method. These issues will surely be 
studied further in future with more data available.
It is too early to say if the ratio $R$ (as 
well as other related ones) will make a big impact on the CKM 
phenomenology, but at any rate it remains a potentially interesting method
to measure the otherwise daunting angle $\gamma$.

\section*{Acknowledgments}
I would like to acknowledge collaboration
with Hrachia Asatrian, Christoph Greub and David
London on many aspects of $B$ physics discussed here.
In particular, David London provided the two updated 
figures shown in this report.  Thanks are due to Manuel 
Drees and the organizing committee of the APCTP workshop in Seoul
for holding an exciting meeting. 
The hospitality of Prof.~H.S. Song at Seoul National University is 
gratefully acknowledged. This work is partially supported by the 
German-Korean scientific exchange programme DFG-446-KOR-113/72/0.


\end{document}